\documentclass[runningheads]{llncs}
\usepackage{booktabs}
\usepackage[T1]{fontenc}
\usepackage{graphicx}
\usepackage{subcaption}

\graphicspath{{./pics/}}
\usepackage{amsmath, amssymb, amsfonts}
\usepackage{algorithm}
\usepackage{float}         
\usepackage{algorithmic}
\usepackage{multirow}
\usepackage{booktabs}
\usepackage{tabularx}
\usepackage{array}
\usepackage[inline]{enumitem}
\usepackage{subcaption} 

\usepackage{url}

\begin{document}
\title{An Efficient Proximity Graph-based Approach to Table Union Search}

%

\author{Yiming Xie\inst{1} \and
Hua Dai\inst{1} \and
Mingfeng Jiang\inst{1} \and
Pengyue Li\inst{2} \and
zhengkai Zhang \inst{1} \and
Bohan Li \inst{3}}
\authorrunning{F. Yiming Xie et al.}
%
\institute{Nanjing University of Posts and Telecommunication, Nanjing, China\\
\email{\{1224045602, daihua, 2024040413, 1024041138\}@njupt.edu.cn} \and
Wuhan University, Wuhan, China\\
\email{pengyueli@whu.edu.cn}\and
Nanjing University of Aeronautics and Astronautics, Nanjing, China\\
\email{bhli@nuaa.edu.cn}\inst{6}
}

%
%
%
\maketitle              

\begin{abstract}
Neural embedding models are extensively employed in the table union search problem, which aims to find semantically compatible tables that can be merged with a given query table. In particular, multi-vector models, which represent a table as a vector set (typically one vector per column), have been demonstrated to achieve superior retrieval quality by capturing fine-grained semantic alignments. However, this problem faces more severe efficiency challenges than the single-vector problem due to the inherent dependency on bipartite graph maximum matching to compute unionability scores. Therefore, this paper proposes an efficient \textbf{P}roximity \textbf{G}raph-based \textbf{T}able \textbf{U}nion \textbf{S}earch (PGTUS) approach. PGTUS employs a multi-stage pipeline that combines a novel refinement strategy, a  filtering strategy based on many-to-one bipartite matching. Besides, we propose an enhanced pruning strategy to prune the candidate set, which further improve the search efficiency. Extensive experiments on six benchmark datasets demonstrate that our approach achieves 3.6-6.0× speedup over existing approaches while maintaining comparable recall rates.


\end{abstract}

\section{Introduction}
\label{sec:intro}
Virtually a significant portion of dataset search tasks emphasize table data~\cite{ts_octus,integrating_data_lake_tables,gent_table_reclamation_in_data_lakes}, as it represents the predominant format for datasets in both web and enterprise environments, including web tables, spreadsheets, CSV files, and database relations. Table Union Search (TUS), retrieving tables that can be meaningfully unioned with a given query table (i.e., those that are compatible in schema and semantically aligned in content), is important for tasks like dataset augmentation and integration.

Recent approaches have transformed TUS into a vector set retrieval problem by representing tables as sets of vectors~\cite{ts_pexeso,ts_starmie,ts_deepjoin}, where each vector corresponds to a column in the table. This vectorized representation captures both structural and semantic information of tables, enabling more sophisticated similarity computations that go beyond simple keyword matching or schema alignment. The vector set retrieval problem has been extensively studied in recent years, and various approaches have been proposed~\cite{mv_search_dessert,mv_search_muvera,mv_search_plaid,mv_search_igp,mv_biovss}. These approaches focus on similarity metrics such as Chamfer distance and Hausdorff distance, which are widely applied in the document search domain. However, existing vector set retrieval methods cannot be applied to table union search, which requires a bipartite assignment between columns. Moreover, the approach proposed by Starmie~\cite{ts_starmie} suffers from high search latency, which renders it impractical for web-scale data lakes with millions of tables.

\textbf{Contributions.} In this paper, we propose an efficient Proximity Graph-based approach to Table Union Search (PGTUS), which which reduces the latency of vector-set based table union search. Our key contributions include:

\begin{itemize}
\item We formulate the vector set-based table union search by converting each column of the table into a vector, and employ the maximum bipartite matching to measure the table unionability.
\item We propose PGTUS by designing a novel refinement and filtering strategy that incorporates many-to-one matching to filter out unpromising candidates.
\item We present an enhanced version of PGTUS called PGTUS$+$, which employs a dual-ended priority queue-based pruning strategy to reduce unnecessary computations.
\item We conduct extensive experiments on real-world datasets to evaluate the performance of PGTUS, demonstrating its effectiveness and efficiency compared to existing methods.
\end{itemize}

\section{Related Work}
\label{sec:related_work}
The integration of table union search and vector set retrieval addresses key challenges in semantic data discovery and efficient similarity matching for large-scale datasets. This section reviews foundational and recent advancements in these areas, highlighting their synergies and limitations that motivate our approach.

\textbf{Table Union Search.} Early work on finding unionable tables used table clustering followed by simple syntactic measures such as the difference in column mean string length and cosine similarities to determine if two tables are unionable~\cite{ts_octus}. Santos~\cite{ts_santos} first use the semantic relationships between pairs of columns in a table to improve the accuracy of table union search. Pylon~\cite{ts_pylon} and Starmie~\cite{ts_starmie} leverages self-supervised contrastive learning to learn a semantic representation of tables, which is then used to find unionable tables.

\textbf{Vector Set Retrieval.} Vector set retrieval has gained attention with the rise of large-scale vector databases. Multi-vector retrieval models (e.g., ColBERT and its variants)~\cite{mv_models_colbert,mv_models_colbertv2} encode images, documents, and other data types into dense vector representations, enabling efficient similarity search and retrieval. Recently, advancements in multi-vector retrieval methods have focused on improving the efficiency and effectiveness of retrieval systems. DESSERT~\cite{mv_search_dessert} first formalized the vector set retrieval problem in a rigorous theoretical framework and proposed a novel retrieval algorithm based on LSH~\cite{lsh_falconn}. PLAID engine~\cite{mv_search_plaid} significantly reduces the retrieval latency of ColBERTv2 through innovative centroid interaction and centroid pruning mechanisms. MUVERA~\cite{mv_search_muvera} reduces multi-vector similarity search to single-vector similarity search, which can be efficiently solved by existing single-vector search methods. IGP~\cite{mv_search_igp} introduces a novel incremental graph traversal technique to facilitate high quality candidate generation. However, the proposed methods often rely on Chamfer distance~\cite{dist_chamfer_linear}, which has limitations in terms of scalability, particularly the lack of theoretical support for maximum bipartite matching~\cite{hungarian_classical}.

\section{Table Unionability Model}
\label{sec:problem_def}

\begin{definition}[Table Repository]
    \label{def:table_repository}
    A table repository $D_T = \lbrace T_{1}, T_{2},
        \ldots, T_{n}\rbrace$ is a collection of tables, where each $T_{i}$ is a table with $|T_i|$ columns.
\end{definition}

We represent each table as a set of column vectors generated by pretrained models, which capture both structural and semantic information of the columns.

\begin{definition}[Vector Set Repository]
    \label{def:vector_set_repository}
    Given a table repository $D_T$, we transform each table $T_i \in D_T$ into a vector set $V_i$ consisting of $|T_i|$ vectors, where each vector in $V_i$ uniquely corresponds to each column in $T_i$. The resulting collection of these vector sets forms a vector set repository denoted as $D_E = \lbrace V_{1}, V_{2}, \ldots, V_{n}\rbrace$.
\end{definition}


\begin{definition}[t-Matching]
Given two tables $T_i,T_j\in D_T$ with their corresponding $V_i,V_j\in D_E$ and a similarity threshold $\tau > 0$, a t-matching is a one-to-one mapping $h_a: V_i' \to V_j'$, where $|V_i'|=t$, $V_i' \subseteq V_i$ and $V_j' \subseteq V_j$, such that for each $v_p \in V_i'$, the pair $(v_p, h_a(v_p))$ satisfies $\langle v_p, h_a(v_p) \rangle \geq \tau$.
\end{definition}



\begin{definition}[Maximum t-Matching]
Given two tables $T_i,T_j\in D_T$ with their corresponding $V_i,V_j\in D_E$ and a similarity threshold $\tau > 0$,a maximum t-matching is the mapping $h_a$ that contains the largest possible number of matched pairs. There may be more than one maximum t-matching, and we denote all maximum t-matchings from $V_i$ to $V_j$ under $\tau$ as $M_a(V_i, V_j, \tau)$.
\end{definition}

\begin{definition}[Table Unionability]
    \label{def:maxmatch}
     Given two tables $T_i,T_j\in D_T$ with their corresponding $V_i,V_j\in D_E$. Let $h_a$ be a maximum t-matching over $V_i'$. We can compute the table unionability between $T_i$ and $T_j$. The table unionability between $T_i$ and $T_j$ is defined as $U(V_i, V_j, \tau)$,
    \[
        U(V_i, V_j, \tau) = \max\limits_{h_a\in M_a(V_i, V_j,\tau)}\sum_{v_p \in V_i'}{\langle v_p, h_a(v_p) \rangle}.
    \]
\end{definition}


\begin{definition}[Table Union Search]
    Given a query table $T_Q$, a table repository $D_T$ and a similarity threshold $\tau$, the goal of table union search is to find the top-$k$ tables in $D_T$ that are most
    unionable with $T_Q$:
    \[
        \mathcal{S}= \operatorname{arg\,max}^{k}_{T_i^*\in D_T}{ U(V_Q, V_{i}^*, \tau)},
    \]
where $V_Q$ and $V_{i}^*$ are the corresponding vector sets of $T_Q$ and $T_i^*$, $\operatorname{arg\,max}^{k}_{T_i^*\in D_T}$ selects $k$ tables $T_i^*$ maximum $U(V_Q, V_i^*, \tau)$ and $T_k^*$ has the $k$-th largest table unionability from $T_Q$.
\end{definition}

\section{Proposed Approach }
\label{sec:proposed_approach}

\subsection{Clustering Based Matching}

In the proposed approach, the table repository $D_T$ is transformed into a vector set repository $D_E$, which incurs substantial memory overhead. To alleviate both the elevated computational demands of table unionability assessment and the associated storage challenges, we employ Vector Quantization (VQ) to compress $D_E$ and introduce a clustering based matching algorithm, which partitions the constituent vectors into compact clusters, upon which the maximum matching algorithm is executed independently, thereby distributing the workload and enhancing scalability.

We first outline the partitioning procedure. Let $E = \{ v_p \mid v_p \in V_i \wedge V_i \in D_E \}$ denote the aggregate set of all constituent vectors across the vector sets in $D_E$. Given a predefined number of clusters $n_c$, we apply $K$-means clustering ($K = n_c$) to $E$, yielding a set of centroid vectors $C_E = \{ c_x\ \}_{x=1}^{n_c}$. For each vector $v \in E$, we assign it to its nearest centroid $\mu(v) \in C_E$ based on Euclidean distance. We then construct an inverted index $I_v = \{ (c_x, F(c_x)) \mid c_x \in C_E \}$, where $F(c_x) = \{ v \mid v \in E \wedge \mu(v) = c_x \}$ maps each centroid to the subset of vectors it governs. Additionally, we build an index $I_w = \{ (i, W(V_i)) \mid V_i \in D_E \}$, where $W(V_i) = \{ (c_x, N)\}$ records, for each $V_i$, the size $N$ of the set of its vectors assigned to each associated centroid $c_x \in L(V_i)$.


\begin{algorithm}[!htb]
\caption{$BuildPartitionInvertedIndex(D_E,\rho_l, \rho_h)$}
\label{alg:centroid_post_process}
\begin{algorithmic}[1]
\renewcommand{\algorithmicrequire}{\textbf{Input:}}
\renewcommand{\algorithmicensure}{\textbf{Output:}}
\REQUIRE $D_E$: vector set repository, dispersion thresholds $\rho_l$, $\rho_h$
\ENSURE $I_p$: inverted index
\STATE $I_p \leftarrow \emptyset$
\FOR{each $V_i \in D_E$}
    \STATE $L(V_i) \leftarrow \{ \mu(v) \mid v \in V_i \}$
    \STATE $\rho(V_i) \leftarrow |L(V_i)| / |V_i|$
    \IF{$\rho(V_i) \leq \rho_l$}
        \STATE Run K-means on $V_i$ to obtain cascade centroids $R$
        \STATE $P_i \leftarrow \{ (G_k, S_k) \mid G_k = \{r_k\}, S_k=\{v \ | \ v\in V_i \wedge v \text{ is assigned to }r_k \}, \, r_k \in C_R \}$
    \ELSIF{$\rho(V_i) \geq \rho_h$}
        \STATE $P_i \leftarrow \{ (G_k, S_k) \mid G_k = \{c_k\}, S_k=\{ v_p \ | \ \mu(v_p)=c_x \wedge v_p \in V_i\}, \, c_x \in L(V_i) \}$
        \WHILE{$\rho(V_i) \geq \rho_h$}
            \STATE $(G_p, S_p), (G_q, S_q) \leftarrow \max\{gsim(G_p,G_q) \ | \ {(G_p, S_p)}, (G_q, S_q) \in P_i\}$
            \STATE $G_{\text{m}} \leftarrow G_p \cup G_q$
            \STATE $S_{\text{m}} \leftarrow S_p \cup S_q$
            \STATE Remove $(G_p, S_p)$ and $(G_q, S_q)$ from $P_i$
            \STATE Add $(G_{\text{m}}, S_{\text{m}})$ to $P_i$
            \STATE $\rho(V_i) \leftarrow |P_i| / |V_i|$
        \ENDWHILE
        \ELSE
        \STATE $P_i \leftarrow \{ (G_k, S_k) \mid G_k = \{c_k\}, S_k \subseteq V_i \text{ assigned to } c_k, \, c_k \in L(V_i) \}$
    \ENDIF
    \STATE $I_p[i] \leftarrow P_i$
\ENDFOR
\RETURN $I_p$
\end{algorithmic}
\end{algorithm}

Although the initial clustering greatly compresses the search space, empirical observations reveal uneven vector distributions across clusters in certain datasets that some vector sets $V_i$ concentrate their vectors around a small number of centroids, whereas others disperse across many. To address this imbalance, we introduce a dispersion metric $\rho(V_i) = |L(V_i)| / |V_i|$, where $L(V_i) = \{ \mu(v_p) \mid v_p \in V_i \}$ captures the unique centroid vectors associated with $V_i$. We start by partitioning each $V_i$ into singleton groups, where each group pairs a centroid $c_p \in L(V_i)$ with the subset of vectors assigned to it. Given thresholds $\rho_l$ and $\rho_h$, we adapt these groups based on $\rho(V_i)$. For highly dispersed sets ($\rho(V_i) \geq \rho_h$), we iteratively merge pairs of groups that maximize the group similarity $\mathrm{gsim}(G_p, G_q) = \max_{c_x \in G_p, c_y \in G_q} \langle c_x, c_y \rangle$ until $\rho(V_j) \leq \rho_h$. For lowly dispersed sets ($\rho(V_i) \leq \rho_l$), we generate additional cascade centroids via K-means on $V_i$ (collected as $C_R$). This yields refined partitions $P_i = \{ (G_k, S_k) \mid k \in [1, m_i] \}$ for each $V_i$, where $G_k \subseteq C_E \cup C_R$ is a non-empty centroid group and $S_k \subseteq V_i$ the assigned vectors (with $\bigcup S_k = V_i$ and disjoint $S_k$), forming the index $I_p = \{ (i, P_i) \mid V_i \in D_E \}$, as detailed in Algorithm~\ref{alg:centroid_post_process}.

Building upon the refined partitions of each vector set $V_i \in D_E$, we seek to establish a many-to-one matching between elements of $V_Q$ and $P_i$, where $V_Q$ is the corresponding vector set of the query table $T_Q$ and $P_i$ is the partitions of $V_i$ denoted by $I_p[i]$.

\begin{definition}[Many-to-One t-Matching]
Given a similarity threshold $\tau > 0$, a many-to-one t-matching is a mapping $h_b$: $V_Q' \to P_i'$, where $|V_Q'|=t$, $V_Q' \subseteq V_Q$ and $P_i' \subseteq P_i$, such that each vector in $V_Q'$ is matched at most once, and each partition $(G_k,S_k) \in P_i'$ is matched at most $|S_k|$ times.
\end{definition}

\begin{definition}[Maximum Many-to-One t-Matching]
A maximum many-to-one t-matching is the mapping $h_b$ that contains the largest possible number of matched pairs. There may be many maximum many-to-one t-matchings and we denote all maximum many-to-one t-matchings from $V_Q \to P_i$ under $\tau$ as $M_b(V_Q, P_i, \tau)$. 
\end{definition}

\begin{definition}[Maximum Weight Many-to-One t-Matching, MWMTO]
    Let $h_b$ be a maximum t-matching over $V_Q'=\{ v_1, v_2,...,v_c\}$ and $h_b(v_q)=(G_k, S_k)$. We can then define the maximum weight one-to-many t-matching. Given a similarity threshold $\tau > 0$, the maximum weight many-to-one t-matching between $V_Q$ and $P_i$ is defined as
    \[
        \text{MWMTO}(V_Q, P_i, \tau) = \mathop{\max}\limits_{h_b\in M_b(V_Q, P_i, \tau)} \sum_{v_q\in V_Q'}{sim(v_q, h_b(v_q))},
    \]
    where $sim(v_q, h_b(v_q))=\max_{c_x\in G_k} \ {\langle v_q, c_x \rangle}$ indicates the similarity between $v_q$ and $G_k$ composed of a set of vectors, and $\langle v_q, c_x \rangle$ is the inner product between vector $v_q$ and $c_x$.
\end{definition}


\begin{figure}[!htb]
    \centering 
    \begin{subfigure}[b]{0.48\textwidth}
        \centering
        \includegraphics[height=6cm]{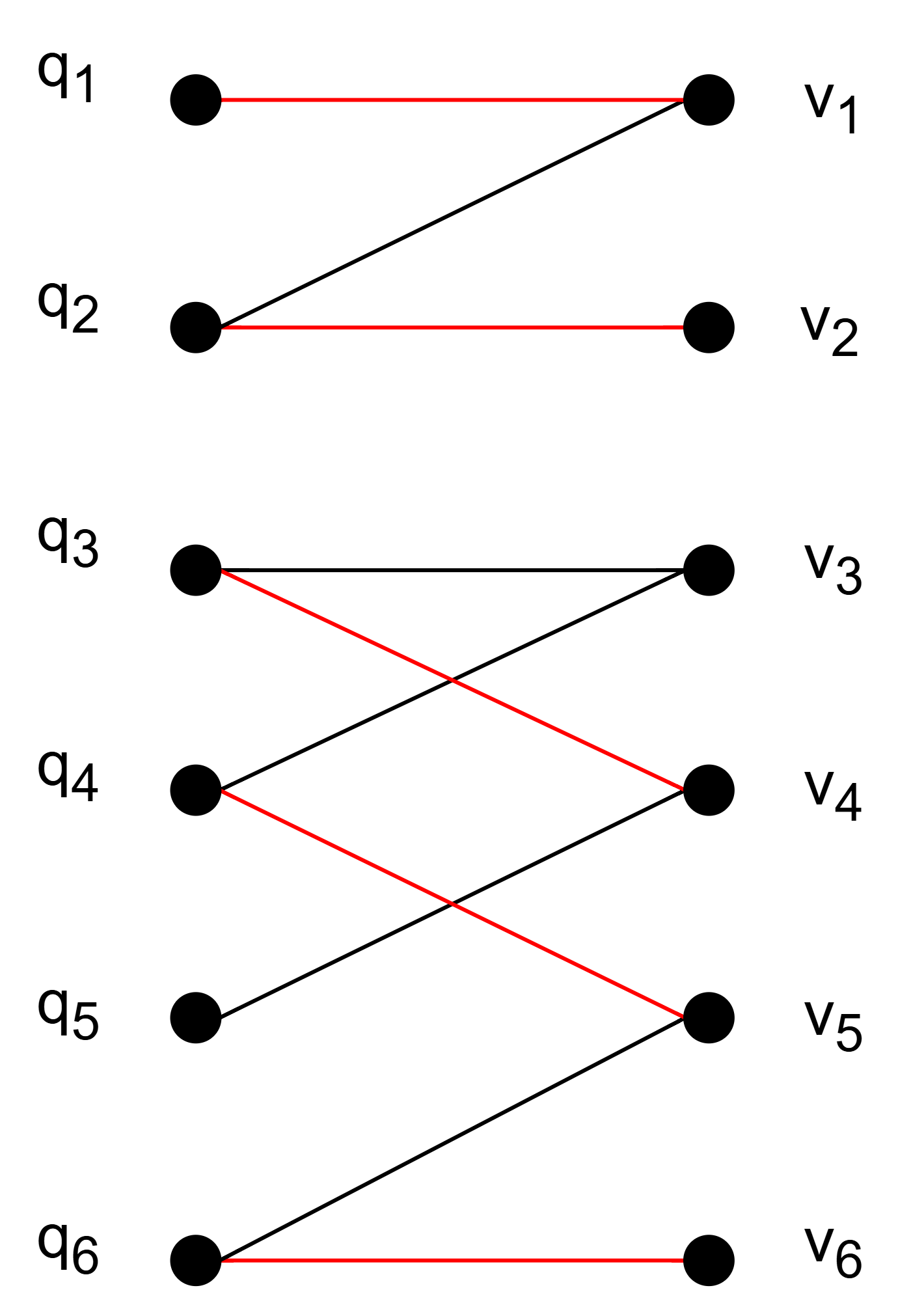}
        \caption{One-to-One t-Matching}
        \label{fig:exp:table_unionability}
    \end{subfigure}
    \hfill
    \begin{subfigure}[b]{0.48\textwidth}
        \centering
        \includegraphics[height=6cm]{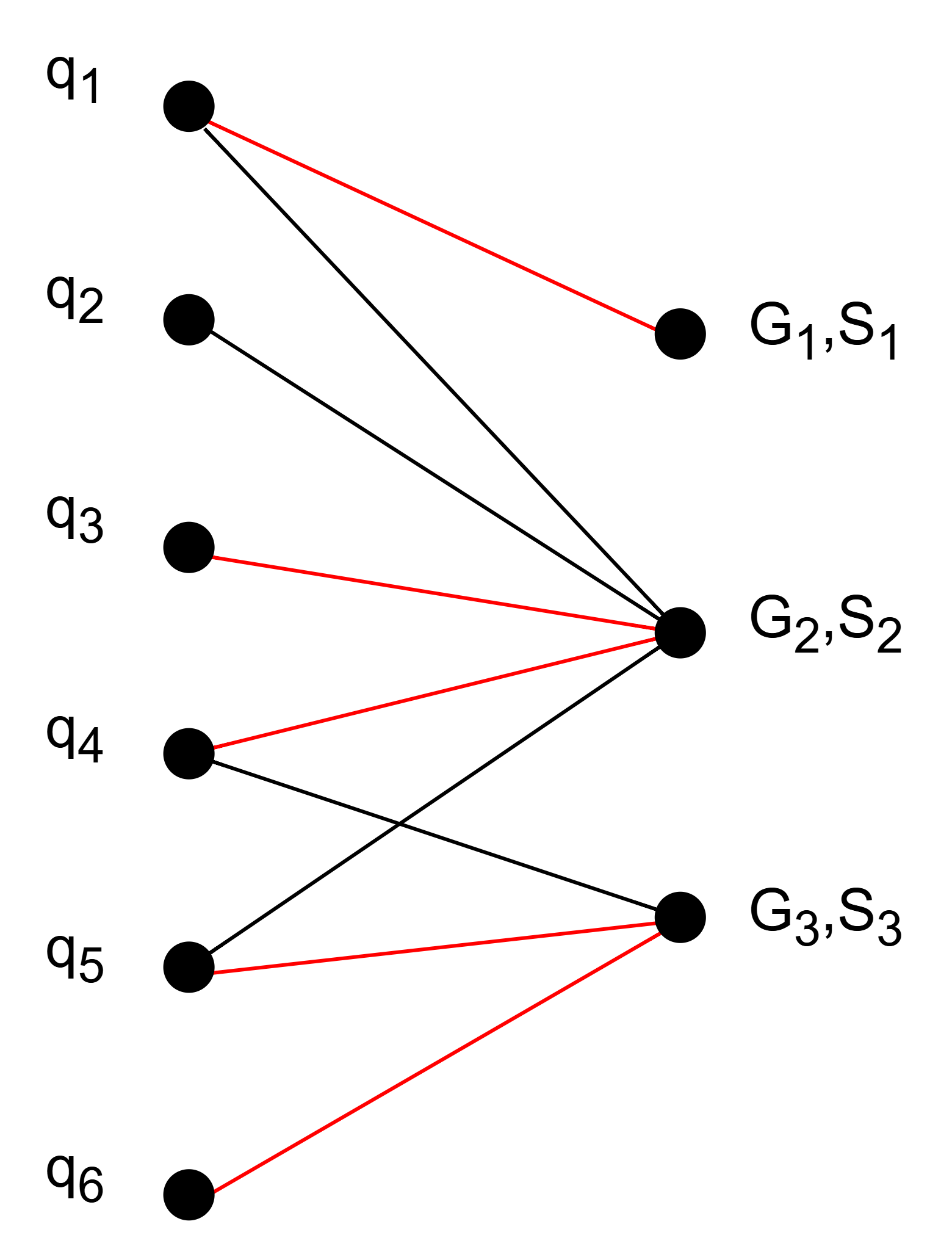}
        \caption{Many-to-One t-Matching}
        \label{fig:exp:MWMTO}
    \end{subfigure}
    \caption{Example for One-to-One and Many-to-One t-Matching.}
    \label{fig:exp:matching}
\end{figure}

\begin{example}
\label{exp:otmam}
Fig.~\ref{fig:exp:matching} presents a comparative illustration of one-to-one and many-to-one maximum matchings. Fig.~\ref{fig:exp:table_unionability} provides an example of table unionability, grounded in one-to-one maximum matching. In contrast, Fig.~\ref{fig:exp:MWMTO} depicts many-to-one maximum matching, where the target vector set $V_i$ is divided into partitions $P_i = \{ (G_1, S_2), (G_2, S_2), (G_3, S_3)\}$ (e.g., $G_2 = ({ c_2, c_3})$, $S_2={ v_2, v_3, v_4}$), and the MWMTO is highlighted in red.
\end{example}


The computational cost of MWMTO is substantially lower than that of full table unionability evaluation, which relies on computationally intensive one-to-one matching algorithms such as the Hungarian algorithm. Consequently, we apply MWMTO between the query vector set $V_Q$ and the partitions of each candidate vector set $V_i \in D_E$ to efficiently identify more promising candidates.

\begin{algorithm}[!htb]
    \caption{$Refinement$($V_Q$, $\phi_c$, $\phi_{ref}$)}
    \label{alg:candidate_refine_pure}
    \begin{algorithmic}[1]
        \renewcommand{\algorithmicrequire}{\textbf{Input:}}
        \renewcommand{\algorithmicensure}{\textbf{Output:}}
        \REQUIRE Query vector set $V_Q$, centroid probe size $\phi_{c}$, reuslt size $\phi_{ref}$
        \ENSURE Top-$\phi_{ref}$ vector sets $R$
        \STATE $H \leftarrow$ Initialize a max-heap storing tuples of the form $(\langle c_j, v_q \rangle, (c_j, v_q))$, ordered by the first component in descending order
        \STATE $\Phi_v, \Phi_u, \Phi_s \leftarrow$ Initialize hash tables with the key type as vector set ID
        \FOR{each query vector $v_q \in V_Q$}
        \STATE $C_{top} \leftarrow$ Find top-$\phi_c$ nearest centroids from $G$ for $v_q$
        \FOR{each centroid $c_j \in C_{top}$}
        \STATE Insert into $H$ the pair with priority $\langle c_j, v_q \rangle$ and payload $(c_j, v_q)$
        \ENDFOR
        \ENDFOR
        \WHILE{$H$ is not empty}
        \STATE $(c_{b}, v_q) \leftarrow$ Get the top entry with highest-score from $H$
        \FOR{V in $I_v[c_b]$}
        \STATE $V_i \leftarrow$ Find the vector set that contains v
        \IF{$i$ not in $\Phi$}
        \STATE $\Phi_s[i]$.score $\leftarrow 0$
        \STATE Initialize $\Phi_v[i][v_q]$ to false for each $v_q \in Q$
        \ENDIF
        \IF{$\Phi_v[i][v_q]$ = false}
        \IF{$\Phi_u[i][c_{b}] < I_w[i][c_{b}]$}
        \STATE $\Phi_u[i][c_{b}] \leftarrow \Phi_u[i][c_{b}] + 1$
        \STATE $\Phi_s[i]\leftarrow \Phi_s[i] + \langle c_{b}, v_q \rangle$
        \ENDIF
        \STATE $\Phi_v[i][v_q] \leftarrow true$
        \ENDIF
        \ENDFOR
        \ENDWHILE
        \RETURN Top-$\phi_{ref}$ vector sets from $\Phi_s$ sorted by score in descending order
    \end{algorithmic}
\end{algorithm}

\subsection{Candidate Refinement}
\label{sec:refinement}

Although the MWMTO provides substantially lower computational cost than one-to-one matching, its complexity remains high for large-scale datasets, necessitating further optimization. Inspired by prior works~\cite{mv_search_plaid,ts_starmie}, which observed that only a small subset of centroids holds high relevance for any given query, we adopt an efficient refinement algorithm over the centroid space. Specifically, given a query table $T_Q$ with its vector set $V_Q$ and the set of centroid vectors $C_E$, we first construct a vector index $G$ (e.g., HNSW) on $C_E$. For each query vector in $V_Q$, we retrieve the top-$\phi_c$ nearest centroids, forming an initial candidate set of associated vector subsets from $D_E$. To further prune low-relevance candidates, we then apply an incremental refinement strategy, yielding the top-$\phi_{ref}$ refined candidates for subsequent filtering.

This strategy enqueues high-similarity centroid-query pairs into a max-heap and iteratively aggregates approximate matching scores across candidate vector sets $V_i$, while enforcing per-centroid capacity constraints. To facilitate efficient candidate pruning, three auxiliary hash tables $\Phi_v$, $\Phi_u$, and $\Phi_s$ are employed, all keyed by vector set IDs. For any candidate $V_i$, $\Phi_v[i][v_q]$ flags whether query vector $v_q$ has processed $V_i$, $\Phi_u[i][c_b]$ monitors $V_i$'s remaining capacity in cluster $c_b$ and $\Phi_s[i]$ stores $V_i$'s cumulative score. In the matching phase, the highest-scoring pair is iteratively extracted from the heap. For each associated candidate set, the procedure checks if the query vector is unvisited and capacity remains; if so, it updates the count and increments the score. The top-$\phi_{ref}$ candidates, ranked by accumulated scores, are returned, as detailed in Algorithm~\ref{alg:candidate_refine_pure}.

\begin{algorithm}[!htb]
    \caption{$BoundsForMWMTO$($V_Q$, $V_i$, $\tau$)}
    \label{alg:mu_bounds}
    \begin{algorithmic}[1]
        \renewcommand{\algorithmicrequire}{\textbf{Input:}}
        \renewcommand{\algorithmicensure}{\textbf{Output:}}
        \REQUIRE Vector set $V_Q$, vector set $V_i \in D_E$, threshold $\tau$
        \ENSURE Lower bound $LB$, upper bound $\!UB$
        \STATE $P_i \leftarrow I_p[i], lb \leftarrow 0$, $ub \leftarrow 0$
        \STATE $U \leftarrow$ Initialize a list of length $|P_i|$
        \FOR{$(G_k, S_k) \in |P_i|$}
            \STATE $U[k] \leftarrow |S_k|$
        \ENDFOR
        \STATE $H \leftarrow \phi$
        \FOR{each $v_q \in V_Q$}
            \STATE $H_q \leftarrow \emptyset$
            \FOR{each $(G_k, S_k) \in P_i$}
                \STATE $sim \leftarrow \max_{c \in G_k} \langle v_q, c \rangle$
                \IF{$sim \geq \tau$}
                    \STATE Insert $(sim, k)$ into $H$
                \ENDIF
            \ENDFOR
            \STATE Sort $H_q$ descending by similarity
            \STATE Insert all elements in $H_q$ to $H$
            \FOR{each $(s, k)$ in $H$}
                \IF{$U[k] > 0$}
                    \STATE $lb \leftarrow lb + s$
                    \STATE $U[k] \leftarrow U[k] - 1$
                    \STATE break
                \ENDIF
            \ENDFOR
        \ENDFOR
        \STATE Sort $H$ descending by similarity;
        \STATE $ub \leftarrow$ sum of the first $\min(|V_Q|, |V_i|)$ similarities in $H$
        \RETURN $lb, ub$
    \end{algorithmic}
\end{algorithm}

\subsection{Candidate Filtering and Scoring}
Following the refinement stage, which yields the top-$\phi_{ref}$ candidate vector sets, we apply MWOTM scoring to further distill the top-$\phi_r$ most promising candidates. These are subsequently ranked according to their unionability scores, derived from the clustering based matching. We extend the pruning strategy in Starmie~\cite{ts_starmie} to MWMTO, enabling efficient pruning of unpromising candidate vector sets while guaranteeing the top-$\phi_r$ results. Specifically, Algorithm~\ref{alg:mu_bounds} derives MWMTO bounds over $P_i$'s partitions by assessing $\max_{c \in G_k} \langle v_q, c \rangle \geq \tau$ for each $v_q \in V_Q$ and $G_k$, gathering valid similarities into a global sorted pool $H$. The lower bound greedily assigns each $v_q$ to the highest-scoring partition with available capacity and accumulates these scores. The upper bound sums the top $\min(|V_Q|, |V_i|)$ similarities from $H$ to provide an optimistic estimate. In the final scoring phase, the bounds of table unionability are transformed into the sum of the lower bounds (upper bounds) for each partition of $V_i$ and top-$k$ results are selected with the highest scores using the same pruning strategy. 

In the final scoring phase, we aggregate the bounds of table unionability over partitions. The upper bound $U\!B_k(V_Q, V_i)$ greedily selects high-weight edges in descending order to estimate an overestimate of the maximum bipartite matching weight between node sets $V_Q$ and $V_i$, aiming for full coverage under matching constraints, while the lower bound $LB_k(V_Q, V_i)$ employs a similar greedy approach to form a partial matching, yielding a conservative weight underestimate since it may not cover all nodes. These bounds are aggregated as $LB = \sum_{k=1}^{|P_i|} LB_k(V_Q, V_i, \tau)$ and $\!UB = \sum_{k=1}^{|P_i|} \!UB_k(V_Q, V_i, \tau)$, from which we select the top-$k$ results with the highest scores using the same pruning strategy. The aggregate score is $\textit{Score}(V_Q, V_i, \tau) = \sum_{k=1}^{|P_i|} U(V_Q^k, S_k, \tau)$.

\begin{algorithm}[!htb]
    \caption{$BasePrune(V_Q, F, \tau, k)$}
    \label{alg:base_prune}
    \begin{algorithmic}[1]
        \renewcommand{\algorithmicrequire}{\textbf{Input:}}
        \renewcommand{\algorithmicensure}{\textbf{Output:}}
        \REQUIRE Query vector set $V_Q$, the set of candidate vector sets $F$, threshold $\tau$, result size $k$
        \ENSURE Top-$k$ vector sets

        \STATE Initialize $H$ as min-heap to store top-$k$ items
        \FOR{each vector set $V_c \in F$}
        \IF{$|H| < k$}
        \STATE $score \leftarrow$ Compute $Score(V_Q, V_i, \tau)$ and add $V_i$ into $H$
        \ELSE
        \IF{$LB(V_Q, V_i, \tau) > H$.top().score}
        \STATE $score \leftarrow$ U$(Q, c, \tau)$
        \STATE Replace the top element of $H$ with $V_c$
        \ELSIF{$ub \geq H$.Top().score}
        \STATE $score \leftarrow$ U$(Q, c, \tau)$
        \IF{$score > H$.Top().score}
        \STATE Replace the top element of $H$ with $V_c$
        \ENDIF
        \ENDIF
        \ENDIF
        \ENDFOR
        \RETURN Top-$k$ vector sets in $H$
    \end{algorithmic}
\end{algorithm}

\section{PGTUS with Enhanced Pruning Strategy}
\label{sec:enhance_pgtus}
Despite the multi-stage search processing and the pruning strategy proposed by Stamie filters out most candidates, PGTUS still needs to deal with a large number of candidates. Therefore, we propose PGTUS$+$, where we utilize the enhanced pruning strategy to filter out more unpromising candidates in the filtering and scoring stage. 



\begin{algorithm}[!htb]
\caption{$EnhancedPrune$($V_Q$, $T$, $\tau$, $\phi_r$)}
\label{alg:late_bounded_prune}
\begin{algorithmic}[1]
\renewcommand{\algorithmicrequire}{\textbf{Input:}}
\renewcommand{\algorithmicensure}{\textbf{Output:}}
\REQUIRE Query vector set $V_Q$, candidate set $T$, similarity threshold $\tau$, result size $\phi_r$
\ENSURE Top-$\phi_r$ ranked candidates with matching scores
\STATE Initialize DEPQs $P_{lb}, P_{ub}$ for bound tracking
\STATE Initialize $H$ as min-heap to store top-$\phi_r$ items
\FOR{each vector set $V_c \in T$}
\STATE $(lb, ub) \leftarrow$ compute bounds for $V_Q$ and $V_c$
\IF {$|P_{lb}|< \phi_r$}
\STATE Insert $V_c$ to $P_{lb},P_{ub}$
\ELSE
\STATE $V_{mlb},V_{mub} \leftarrow P_{lb}.Min(),P_{ub}.Min()$
\IF {$ub < V_{mub}.ub$}
\STATE Discard candidate $V_c$
\ELSIF {$lb > V_{mlb}.lb$}
\STATE Remove $V_{mub}$ from both $P_{lb}$ and $P_{ub}$
\STATE Insert $V_c$ to $P_{lb},P_{ub}$
\ELSE
\WHILE{$|P_{ub}|>0$ \textbf{and} $ub \geq V_{mlb}.lb$ \textbf{and} $lb \leq V_{mub}.ub$}
\STATE $V \leftarrow$ The item with maximum upper bound from $P_{ub}$
\STATE Process $V$ as in $BasePrune$
\STATE Break if $|H| \geq \phi_r$ and $V_{c}.ub \leq H.Top().score$
\STATE Remove $V$ from both queues
\STATE $V_{mlb},V_{mub} \leftarrow P_{lb}.Min(), P_{ub}.Min()$
\ENDWHILE
\STATE Insert $V_c$ to $P_{lb},P_{ub}$
\ENDIF
\ENDIF
\ENDFOR
\STATE Process remaining elements using $BasePrune$ in descending order of upper bounds
\RETURN Top-$\phi_r$ candidates in $H$
\end{algorithmic}
\end{algorithm}

Starmie's pruning strategy eliminates most candidate items via bound-based filtering, as described in Algorithm~\ref{alg:base_prune}. Despite its effectiveness, this method suffers from instability in pruning guarantees, particularly with unordered candidate sets, where performance can vary significantly. In the worst case, the min-heap $H$ must be repeatedly updated as candidates with disparate bound qualities arrive, incurring erratic computational costs and potentially undermining overall efficiency.

Motivated by these shortcomings, we propose an enhanced pruning strategy grounded in a fundamental theoretical observation. Consider a candidate set $F = \{V_1, V_2, \ldots, V_n\}$ with associated lower bounds $\{lb_i\}_{i=1}^n$ and upper bounds $\{ub_i\}_{i=1}^n$. For any $V_c \in T$, if at least $\phi_r$ other candidates $V_j \in T \setminus {V_c}$ satisfy $lb_j > ub_c$, then $V_c$ cannot rank among the top-$\phi_r$ highest-scoring items. Our approach maintains two double-ended priority queues (DEPQs), $P_{lb}$ and $P_{ub}$, to track lower and upper bounds, respectively, alongside a min-heap $H$ for final scores. As detailed in Algorithm~\ref{alg:late_bounded_prune}, for each candidate vector set $V_c$, we compute its lower and upper bounds relative to the query set $V_Q$ (line 4). If $|P_{lb}| < \phi_r$, $V_c$ is inserted into both queues (lines 6--8). Otherwise, we extract the candidates with the minimum lower bound ($V_{mlb}$) and minimum upper bound ($V_{mub}$) from $P_{lb}$ and $P_{ub}$, respectively (line 9). A three-way decision then guides the process: (1) If $ub$ of $V_c$ falls below the minimum upper bound in $P_{ub}$, discard $V_c$ outright (lines 11--12); (2) If $lb$ of $V_c$ exceeds the minimum lower bound in $P_{lb}$, replace $V_{mub}$ (the candidate with the weakest upper bound) with $V_c$ in both queues (lines 13--15); (3) Otherwise, process candidates from $P_{ub}$ in descending order of upper bounds until a resolution is reached (lines 17--27).

To efficiently support the bidirectional extreme-value retrieval operations required by our enhanced pruning strategy, we implement a DEPQ data structure.
This DEPQ employs a dual-heap architecture, comprising a binary min-heap $P_{lb}$ and a binary max-heap $P_{ub}$, augmented by a hash table to track element positions. In Algorithm~\ref{alg:late_bounded_prune}, elements must be synchronously removed from both heaps. Extracting the maximum element from $P_{ub}$ requires $O(\log n)$ time, after which the corresponding element must also be removed from $P_{lb}$. However, removing an arbitrary element from $P_{lb}$ first requires locating its position, which can be prohibitively expensive.
To mitigate this, we introduce lazy deletion semantics via tombstone marking. Specifically, when an element requires removal from both heaps, we immediately execute the $O(\log n)$-time removal on the heap where its position is known (i.e., the root of $P_{ub}$), while merely marking it as deleted in $P_{lb}$. The actual removal from $P_{lb}$ is thus deferred until the element naturally percolates to the root during subsequent heap operations, thereby amortizing the removal cost across multiple extractions. This technique preserves $O(\log n)$-time complexity for both insertions and extractions, while enabling efficient synchronized deletions.


\section{Experimental Evaluation}
\label{sec:experiment}

\subsection{Settings}
\textit{1)Datasets:}
We evaluate our method on six data lakes, and table~\ref{tab:dataset_stats} provides their statistics. Specially, the Canada, US, and UK Open Data dataset is originally designed for Table Join Search tasks, where most queries contain only single columns. To adapt it for our Table Union Search evaluation, we selected 1000 data tables with more than 15 columns as query tables, ensuring sufficient structural complexity for meaningful union operations.

\begin{table*}
    \centering
    \caption{Statistics of the datasets used in experiments}
    \begin{tabularx}{\textwidth}{l*{5}{>{\centering\arraybackslash}X}}
        \toprule
        \multirow{2}{*}{\textbf{Dataset}} & \multicolumn{2}{c}{\textbf{Data Lake Tables}} & \multicolumn{2}{c}{\textbf{Query Tables}} \\
        \cmidrule(lr){2-3} \cmidrule(lr){4-5}
                                           & \textbf{\#Tables} & \textbf{\#Columns} & \textbf{\#Queries} & \textbf{\#Avg Cols} \\
        \midrule
        Santos Small~\cite{ts_santos}                 & 550   & 6322   & 50    & 12.30 \\
        Santos Large~\cite{ts_santos}                 & 11086 & 121796 & 80    & 12.71 \\
        TUS Small~\cite{ts_table_union_search}        & 1530  & 14810  & 10    & 34.5 \\
        TUS Large~\cite{ts_table_union_search}        & 5043  & 55K    & 4293  & 11.85 \\
        Open Data~\cite{tjs_josie} & 13K   & 218K   & 1000  & 30.46 \\
        Lakebench Webtable~\cite{deng2024lakebench}   & 2.8M  & 14.8M  & 1000  & 14.57 \\
        \bottomrule
    \end{tabularx}
    \label{tab:dataset_stats}
\end{table*}

\textit{2)Baselines:}
We compare our proposed method against Stamie, the state-of-the-art method on table union search using HNSW index and Stamie (VQ), combining Starmie with vector quantization for vector compression and improved query speed. We also conduct experiments on Stamie$+$ and Starmie (VQ)$+$, which combines our proposed pruning strategy to enhance efficiency through effective pruning.

\textit{3)Evaluation Metrics:}
We evaluate methods using Recall, Latency, and Memory consumption to capture both effectiveness and efficiency. Given a query table $T_q$, let $\mathcal{T}_q$ denote the retrieved set, and let $\mathcal{T}_g$ denote the set of ground-truth unionable tables. Recall is defined as $Recall = \frac{|\mathcal{T}_g \cap \mathcal{T}_q|}{|\mathcal{T}_g|}$. Latency measures the average per-query processing time. For each method, we report the average Recall and Latency. Our ground truth $\mathcal{T}_g$ is established using the Hungarian algorithm to compute exact table unionability scores for all tables in the repository.

\textit{4)Parameter settings:} 
For the top-$k$ table union search, we set $k=50$ for WebTable, OpenData, $k=20$ for Santos Large and $k=10$ for Santos Small, TUS datasets. For candidate generation in Starmie, we tune $k'$ in \{$1k$, $5k$, $10k$, $12k$, $15k$, $20k$, $25k$, $30k$\}. For centroid probe in Starmie(VQ) and PGTUS, we tune $\phi_c$ in \{1, 2, 4, 8, 16, 32, 64\}. The filtering parameter $\phi_{r}$ is uniformly set to $3k$ across all datasets to maintain consistency in the filtering stage. For PGTUS, we tune the refinement parameter $\phi_{ref}$ in \{$k$, $2k$, $3k$, $4k$, $5k$, $8k$, $10k$, $20k$\}. The dimension of embedding generated by Starmie is 768. For HNSW index we set the number of neighbors of each node as 16 and $efConstruction$ as 200.


\textit{5)Implementation details:}
Our experiments were conducted on an Ubuntu system equipped with 32 vCPUs and 512GiB DDR4 RAM. Vector quantization components were implemented in C$++$ for optimal performance, while all other algorithms were implemented in Python. We utilized the HNSW index provided by Milvus vector database and reimplemented Starmie in Python for fair comparison. Our proposed approach employs Milvus vector database deployed via Docker for efficient vector storage and retrieval, and utilizes the K-means clustering algorithm from the FAISS library to perform vector clustering. The same environment and settings were used for all methods to ensure a fair comparison.

\begin{table*}
    \caption{Search latency evaluation of different pruning strategies (s)}
    \resizebox{\linewidth}{!}{
    \begin{tabular}{lcccccc}
        \toprule
        \multirow{2}{*}{\textbf{Method}} & \multicolumn{6}{c}{\textbf{Average Search Latency (s)}}                                                                                                            \\
        \cmidrule(lr){2-7}
                                         & \textbf{Santos Small}                                   & \textbf{Santos Large} & \textbf{TUS Small} & \textbf{TUS Large} & \textbf{Open Data} & \textbf{Webtable} \\
        \midrule
        BF                               & 5.6                                                     & 104.2                 & 61.1               & 121.1              & 37.8               & 4785              \\
        BBP                              & 1.0                                                     & 9.2                   & 8.0                & 4.5                & 12.2               & 233               \\
        EBBP                             & 0.4                                                     & 3.0                   & 1.6                & 0.9                & 6.7                & 176               \\
        \midrule
        \textbf{Speedup (BF)}            & \textbf{15.6×}                                          & \textbf{35.3×}        & \textbf{38.5×}     & \textbf{134.6×}    & \textbf{5.7×}      & \textbf{27.2×}    \\
        \textbf{Speedup (BBP)}          & \textbf{2.9×}                                           & \textbf{3.1×}         & \textbf{5.1×}      & \textbf{4.9×}      & \textbf{1.8×}      & \textbf{1.3×}     \\
        \bottomrule
    \end{tabular}
    \label{tab:pruning_efficiency}
    }
\end{table*}

\subsection{Performance Evaluation and Parameter Analysis}

In this section, we first conduct a comprehensive performance evaluation of our proposed method against baselines across six datasets, demonstrating its efficiency. Subsequently, we perform an in-depth analysis of the impact of key parameters on the performance of our method, providing insights into optimal configurations for various scenarios.

\textit{1) Performance Evaluation :}

\begin{figure*}[ht]
    \centering
    \includegraphics[width=1.0\textwidth]{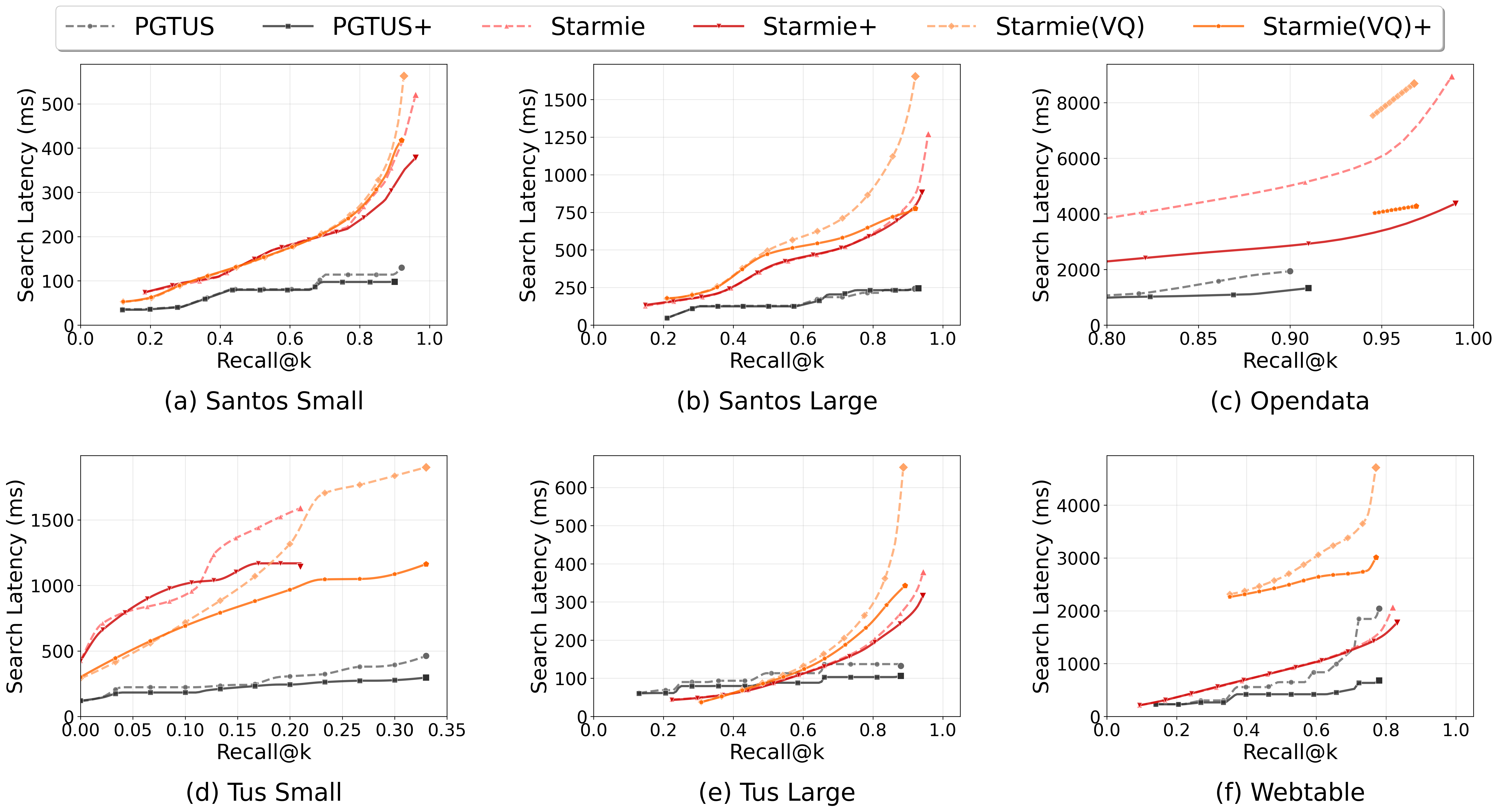}
    \caption{Latency-recall performance evaluation across different datasets.}
    \label{fig:performance_comparison}
\end{figure*}

\textbf{Search Evaluation.} 
Figure~\ref{fig:performance_comparison} shows the latency-recall performance evaluation across six datasets. We use grid search with PGTUS to explore the Pareto frontier. PGTUS offers 2X-4X speedup over Starmie at the same recall level, due to fewer candidates and faster ranking via partitioned computation. The enhanced pruning strategy becomes more effective with more candidates, outperforming Starmie’s strategy by reducing unnecessary computations, especially in large-scale scenarios. The effect is notable on OpenData, where high vector dispersion allows high recall with efficient computation from 1-2 clusters.

\begin{figure}[!htb]
    \centering
    \includegraphics[width=0.8\textwidth]{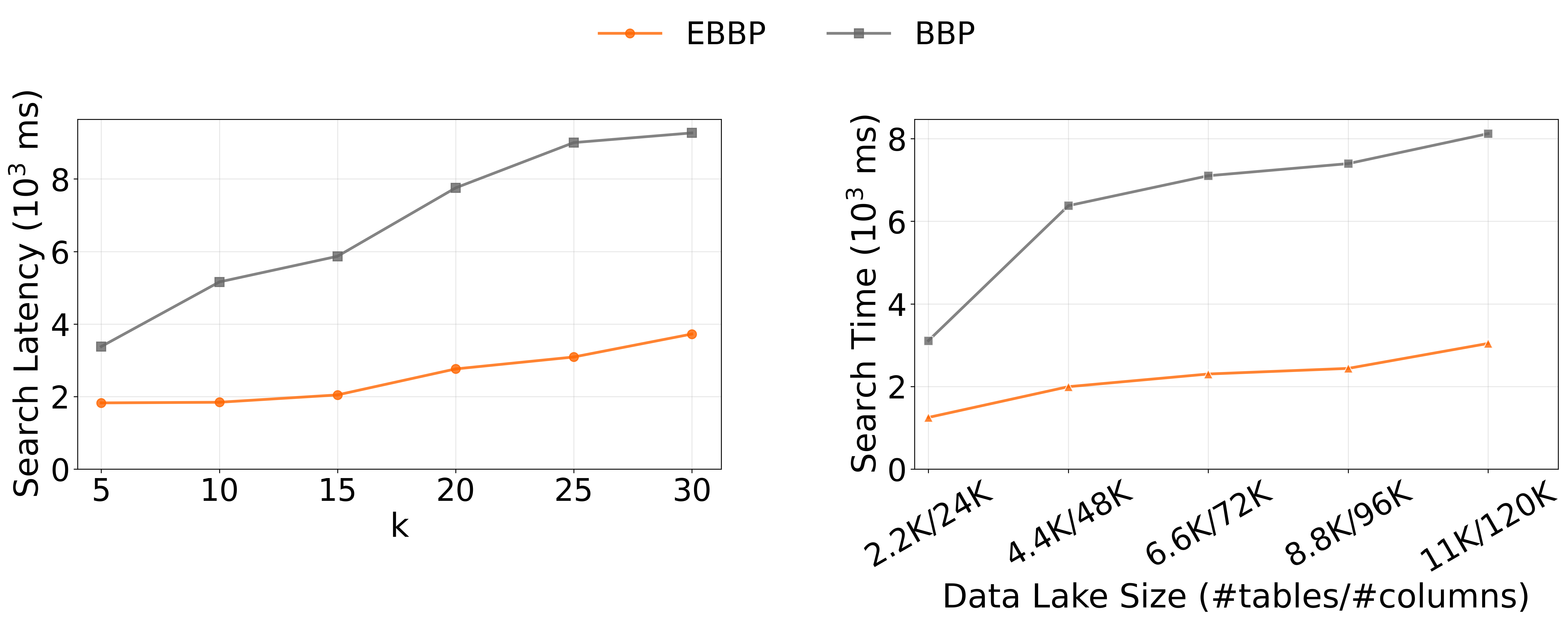}
    \caption{Scalability on the Santos Large benchmark}
    \label{fig:santos_scalability}
\end{figure}

\textbf{Pruning Strategy Evaluation.}
We systematically compare pruning strategies across all datasets via iterative processing, evaluating brute-force exhaustive search (BF) as a baseline, Starmie's strategy (BBP), and our enhanced strategy (EBBP). For the brute-force baseline, we implement exhaustive candidate evaluation without any pruning optimization to establish the computational upper bound. Table~\ref{tab:pruning_efficiency} compares search latency across pruning strategies, demonstrating our method's computational advantages while maintaining retrieval quality. Figure~\ref{fig:santos_scalability} further evaluates scalability on Santos Large, showing our pruning strategy consistently outperforms Starmie's across varying $k$ (5-30) and data lake sizes (1K-11K tables). Our method reduces distance computations by 69\% (53 vs. 169), demonstrating superior efficiency.

\textit{2) Parameter Analysis:}

\begin{figure}[!htb]
    \centering
    \includegraphics[width=1.0\textwidth]{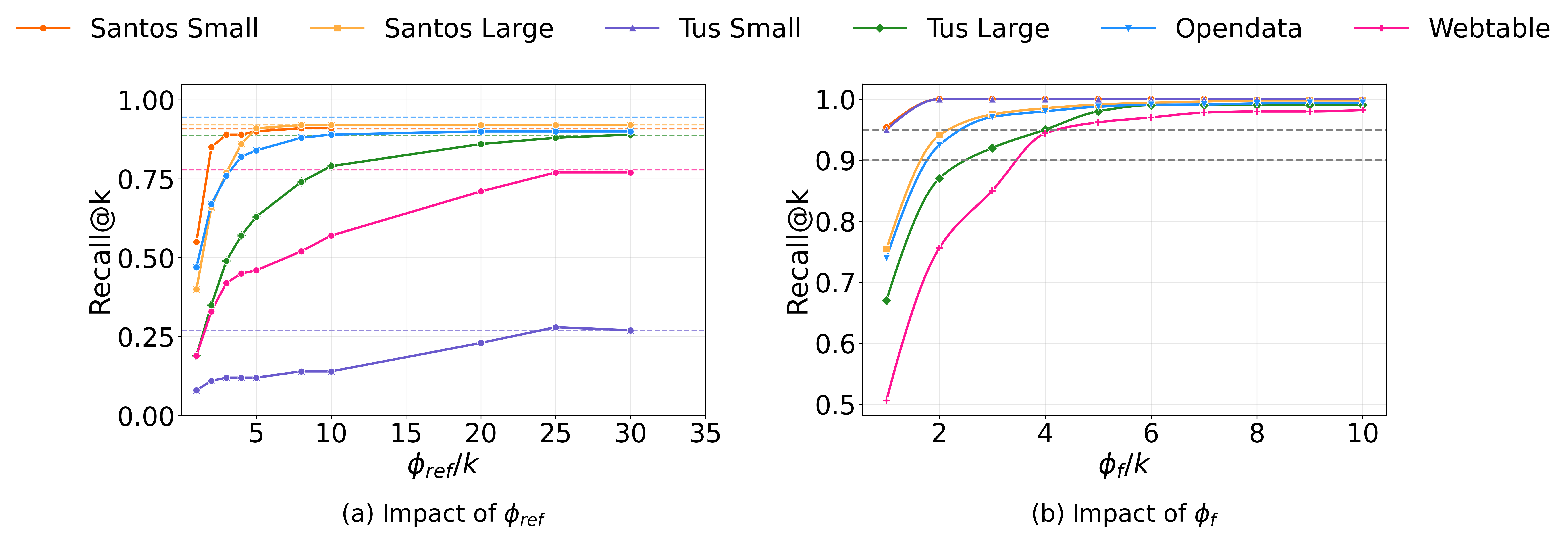}
    \caption{Impact of parameters $\phi_{ref}$, $\phi_{r}$ on recall performance.}
    \label{fig:refinement_and_filter_analysis}
\end{figure}

\textbf{Number of candidates $\phi_{ref}$ in refinement stage.} Figure~\ref{fig:refinement_and_filter_analysis}a presents the relationship between the refinement parameter $\phi_{ref}$ and recall performance across benchmark datasets, with $\phi_c$ fixed at 1 for Open Data and 32 for others. The results demonstrate a consistent trend: recall improves monotonically with the $\phi_{ref}/k$ ratio, but with diminishing marginal returns beyond a dataset-specific threshold. The dashed lines represent Starmie(VQ) performance under identical configurations, providing a comparative baseline.

\textbf{Number of candidates $\phi_{r}$ in filtering stage.} Figure~\ref{fig:refinement_and_filter_analysis}b shows the impact of $\phi_{r}$ on recall across datasets. We conducted iterative searches per dataset, ranking by OTMoM score and measuring recall at varying pool sizes. Recall rises with $\phi_{r}$, especially for small datasets, but gains diminish beyond $\phi_{r} \approx 2k$, balancing precision against latency.

\subsection{Index Construction Evaluation}


Table~\ref{tab:index_storage} reports memory usage and construction time on the largest datasets: Open Data (13K tables, 218K columns) and Webtable (2.8M tables, 14.8M columns). Both PGTUS and Starmie (VQ) achieves a substantial reduction in memory footprint compared to Starmie, primarily through vector quantization. Compared to Starmie (VQ), PGTUS additional stores the inverted index $I_p$ recording vector set partitions.

\begin{table}[ht]
    \centering
    \caption{Index storage and construction time evaluation}
    \begin{tabularx}{\linewidth}{l*{4}{>{\centering\arraybackslash}X}}
        \toprule
        \multirow{2}{*}{\textbf{Method}} & \multicolumn{2}{c}{\textbf{Open Data}} & \multicolumn{2}{c}{\textbf{Webtable}} \\
        \cmidrule(lr){2-3} \cmidrule(lr){4-5}
                                         & \textbf{Storage} & \textbf{Time} & \textbf{Storage} & \textbf{Time} \\
        \midrule
        Starmie          & 1285.85 MB & 5 m   & 98 GB   & 2.5 h \\
        Starmie(VQ)      & 361.29 MB  & 12 s  & 21.8 GB & 1 h   \\
        PGTUS              & 413.71 MB  & 35 s  & 30.6 GB & 2 h   \\
        \bottomrule
    \end{tabularx}
    \label{tab:index_storage}
\end{table}

\section{Conclusion}

In this paper, we proposed an efficient proximity graph-based approach to table union search. We introduced PGTUS and PGTUS$+$, composed of refinement and filtering strategies effectively prunes unpromising candidates, reducing computational overhead. Our dual-ended priority queue-based pruning strategy effectively filter out unpromising candidates, which reduces computational overhead. Empirical validation across six benchmarks show that the proposed approaches achieve 3.6-6.0x speedups over existing approaches.

%
%
%
\bibliographystyle{splncs04}
\bibliography{references}

\end{document}